\documentclass[11pt]{article}

\usepackage{amsmath}
\usepackage{authblk}
\usepackage{enumitem}  
\usepackage{hyperref}  
\usepackage[usenames, dvipsnames]{color} 
\usepackage{graphicx}     
\usepackage{subcaption}
\usepackage[utf8]{inputenc}
\usepackage{url}
\usepackage{cite}
\usepackage{amssymb}
\usepackage{mathrsfs}

\usepackage[dvipsnames]{xcolor}
\usepackage{physics}

\usepackage{overpic}

\usepackage{hyperref}

\hypersetup{
    unicode=false,          % non-Latin characters 
    pdftoolbar=true,        % show Acrobat
    pdfmenubar=true,        % show Acrobat 
    pdffitwindow=false,     % window fit to page when opened  
    pdfstartview={FitH},    % fits the width of the page to the window
    pdfauthor={Author},     % author
    pdfsubject={Subject},   % subject of the document  
    pdfcreator={Creator},   % creator of the document
    pdfproducer={Producer}, % producer of the document
    pdfkeywords={keyword1} {key2} {key3}, % list off keywords
    pdfnewwindow=true,      % links in new window
    linktocpage=true,
    linkcolor=blue,           % color of internal links (change box color with linkbordercolor)
     linkbordercolor=blue,
    citecolor=blue,        % color of links to bibliography
    filecolor=blue,      % color of file links
    urlcolor=blue         % color of external links
} 

\def\beq{\begin{equation}}
\def\eeq{\end{equation}}

\def\d{\delta}
\def\e{\epsilon}
\def\k{\kappa}

\def\o{\omega}
\def\O{\Omega}
\def\L{\Lambda}
\def\S{\Sigma}

\newcommand{\ben}{\begin{enumerate}}
\newcommand{\een}{\end{enumerate}}
\newcommand{\be}{\begin{equation}}
\newcommand{\ee}{\end{equation}}
\newcommand{\mc}{\mathcal}

\newcommand{\diff}{\mathrm{d}}

\usepackage{tikz}
\usetikzlibrary{positioning}
\usetikzlibrary{intersections}
\usetikzlibrary{fadings} 
\usetikzlibrary{arrows.meta} 
\usetikzlibrary{arrows}

\tikzfading[name=fade out,
inner color=transparent!0,
outer color=transparent!100]

\definecolor{cherryblossompink}{rgb}{1.0, 0.72, 0.77}
\definecolor{lightblue}{rgb}{0.68, 0.85, 0.9}

\usetikzlibrary{decorations.pathmorphing}
\usetikzlibrary{decorations.pathreplacing,decorations.markings}

\usetikzlibrary{backgrounds,automata}

\DeclareSymbolFont{matha}{OML}{txmi}{m}{it}% txfonts
\DeclareMathSymbol{v}{\mathord}{matha}{118}

\begin{document}
 
\numberwithin{equation}{section}
 
 \title{\vspace{-2cm}\bf\LARGE The entropy of dynamical de Sitter horizons}

\author{Jinan Zhao\thanks{jinanzhao@mail.bnu.edu.cn}}
\affil{\it{\small School of Physics and Astronomy, Beijing Normal University, Beijing 100875, China}}
%\affil{\it{\small Key Laboratory of Multiscale Spin Physics, Ministry of Education, Beijing Normal University, Beijing 100875, China}}

%\author{\large Jinan Zhao\thanks{jinanzhao@mail.bnu.edu.cn}}

\vspace{10mm}

%\affil{\it{\small School of Physics and Astronomy, Beijing Normal University, Beijing 100875, China}} 

\date{\today}
 
\maketitle

\begin{abstract}
\noindent 

Recently Hollands, Wald and Zhang proposed a new formula for the entropy of a dynamical black hole. We lift this construction to the dynamical cosmological event horizon of an asymptotically de Sitter spacetime. By introducing a nontrivial correction term in the formula for the entropy, we generalize Gibbons and Hawking's ``first law of event horizons" to non-stationary eras. We also develop the non-stationary physical process first law between two arbitrary horizon cross-sections for the cosmological event horizon.
  
\end{abstract}

\thispagestyle{empty}

\newpage

\tableofcontents
 
\section{Introduction} \label{sec:intro}

The discovery of black hole thermodynamics is one of the most remarkable achievements of modern physics\cite{Bekenstein:1973ur,Bardeen:1973gs,Hawking:1975vcx}. However, in the standard treatments of the black hole thermodynamics, the first law often does not hold for non-stationary perturbations of a stationary black hole, and if it does, the entropy cannot be evaluated at an arbitrary horizon cross-section of the perturbed non-stationary black hole\cite{Visser:2024pwz}. Recently, Hollands, Wald and Zhang proposed a new definition for the entropy of a dynamical black hole\cite{Hollands:2024vbe}. By introducing a dynamical correction term to the usual Noether charge formula, they overcame the two limitations above, and established the non-stationary first law for arbitrary horizon cross-sections of a perturbed black hole.

Shortly after the discovery of black hole thermodynamics, Gibbons and Hawking established the thermodynamics for the cosmological event horizons of de Sitter spacetimes \cite{Gibbons:1977mu}. One of the key results they found is the ``first law of event horizons", which states that the variation away from a Kerr-de Sitter black hole spacetime satisfies
\begin{equation}\label{GHfirstlaw}
    \int_\Sigma \delta T_{ab}\xi^a d\Sigma^b = -\kappa_C \delta A_C/8\pi G 
    -\kappa_H \delta A_H/8\pi G - \Omega_H\delta J_H. 
\end{equation}
Here the integral is over a spatial slice $\Sigma$ bounded by the cosmological and black hole horizons, $T_{ab}$ is the matter energy momentum tensor, $\xi^a$ is the 
Killing vector generating the cosmological horizon, the subscripts $C$ and $H$ on the (positive) surface gravities $\kappa$ and areas $A$ refer to the cosmological and black hole horizons respectively, and $\Omega_H$ and $J_H$ are the angular velocity and angular momentum of the black hole relative to the cosmological horizon. However, in the original derivation of the first law, it was assumed that the perturbation performed on the metric is stationary\cite{Gibbons:1977mu}, and in recent studies of the first law, the entropy of the de Sitter horizon is only evaluated at the bifurcation surface of the cosmological horizon\cite{Dolan:2013ft,Jacobson:2018ahi,Banihashemi:2022htw}. Therefore in this paper we would like to extend Gibbons and Hawking's “first law of event horizons” to non-stationary eras, and introduce the new definition for the entropy valid to an arbitrary horizon cross-section. We also would like to develop the “local physical process version” of the first law for the cosmological event horizon.

The rest of this paper is organized as follows: In Sec. \ref{sec2} we review the covariant phase space formalism as well as the definition to the entropy of a dynamical black hole. In Sec. \ref{sec3} we introduce the background geometry of the stationary asymptotically de Sitter black hole spacetime and impose gauge conditions on non-stationary perturbations. In Sec. \ref{sec4} we derive both the non-stationary comparison first law and the non-stationary physical process first law for de Sitter horizons. We end with a summary of results and a discussion of possible future research directions in Sec. \ref{sec5}. 

We will mainly follow the notation and conventions of \cite{Wald:1984rg}. In particular, we use boldface letters to denote differential forms with the tensor indices suppressed. Throughout this paper, we set $c = \hbar = k_B = 1$ while keep Newton’s constant $G$ explicit.

%%%%%%%%%%%%%%%%%%%%%%%%%%%%%%%%%%%%%%%%%%%%%%%%%%%%%%%%%

\section{Review of the dynamical black hole entropy} \label{sec2}

In this section we review the covariant phase space formalism and the strategy to the definition of the dynamical black hole entropy.

Consider an arbitrary diffeomorphism covariant theory of gravity in $n$-dimensions described by a Lagrangian $n$-form $\boldsymbol{L}$. Under a first-order variation of the dynamical fields, the variation of the Lagrangian can always be expressed as
\begin{equation}\label{variation of Lagrangian}
    \delta \boldsymbol{L} = \boldsymbol{E} \delta \phi + \mathrm{d} \delta \boldsymbol{\theta},
\end{equation}
where $\phi$ is the collection of dynamical fields such as metric $g_{ab}$ and other matter fields, $\boldsymbol{E}$ is the equation of motion locally constructed out of $\phi$, and the symplectic potential $(n-1)$-form $\boldsymbol{\theta}(\phi,\delta\phi)$ is locally constructed out of $\phi$, $\delta\phi$ and their derivatives and is linear in $\delta\phi$. The symplectic current $(n-1)$-form $\boldsymbol{\omega}$ is obtained from $\boldsymbol{\theta}$ via
\begin{equation}\label{symplectic current}
    \boldsymbol{\omega} (\phi; \delta_1 \phi, \delta_2 \phi)=\delta_1 \boldsymbol{\theta}(\phi,\delta_2 \phi)-\delta_2 \boldsymbol{\theta}(\phi,\delta_1 \phi).
\end{equation}
Let $\chi^a$ be an arbitrary vector field which is also the infinitesimal generator of a diffeomorphism, then the associated Noether current $(n-1)$-form $\boldsymbol{J}$ is defined by
\begin{equation}\label{Noether current}
    \boldsymbol{J} (\phi) = \boldsymbol{\theta} (\phi, \mathcal{L}_\chi \phi) - \chi \cdot \boldsymbol{L} (\phi),
\end{equation}
where the notation $\cdot$ denotes the contraction of a vector field with the first index of a differential form. It was shown that the Noether current can also be written in the form\cite{Iyer:1995kg,Seifert:2006kv}
\begin{equation}\label{Noether charge}
    \boldsymbol{J} = \mathrm{d}\boldsymbol{Q}[\chi] + \chi^a \boldsymbol{C}_a.
\end{equation}
Where the $(n-2)$-form $\boldsymbol{Q}$ is referred to as the “Noether charge”\cite{Iyer:1994ys} and the dual vector valued $(n-1)$-form $\boldsymbol{C}_a$ vanishes when the equations of motion are satisfied. We vary (\ref{Noether current}) (where the vector field $\chi^a$ is taken to be fixed under the variation) and use (\ref{variation of Lagrangian}) and (\ref{symplectic current}), then we obtain
\begin{equation}
    \delta \boldsymbol{J}(\phi) = -\chi \cdot \left[\boldsymbol{E}(\phi) \delta \phi\right] + \boldsymbol{\omega} (\phi; \delta\phi, \mathcal{L}_\chi \phi)+\mathrm{d}\left[\chi \cdot \boldsymbol{\theta}(\phi,\delta \phi)\right].
\end{equation}
Next we calculate the variation of (\ref{Noether charge}) to obtain the fundamental identity\cite{Hollands:2012sf}
\begin{equation}\label{fundamental identity}
    \boldsymbol{\omega}(\phi; \delta\phi, \mathcal{L}_\chi \phi) = \chi \cdot \left[\boldsymbol{E}(\phi) \delta \phi\right] +\chi^a \delta \boldsymbol{C}_a (\phi)+ \mathrm{d}\left[\delta\boldsymbol{Q}[\chi]-\chi \cdot \boldsymbol{\theta}(\phi,\delta \phi)\right].
\end{equation}
For a stationary black hole with horizon Killing field $\chi^a = \xi^a$, we have $\boldsymbol{\omega}(\phi; \delta\phi, \mathcal{L}_\xi \phi)=0$ as it linearly depends on $\mathcal{L}_\xi \phi$. Thus the fundamental identity becomes
\begin{equation}\label{stationary background}
    \diff (\delta \boldsymbol{Q}[\xi] - \xi \cdot \boldsymbol{\theta}(\phi,\delta \phi)) = -\xi \cdot \left[\boldsymbol{E}(\phi) \delta \phi\right] - \xi^a \delta \boldsymbol{C}_a (\phi).
\end{equation}
When the background field equations and the linearized constraint equations for perturbed fields are satisfied, i.e. $\boldsymbol{E}(\phi) = 0$ and $\delta \boldsymbol{C}_a (\phi) = 0$, the fundamental identity (\ref{stationary background}) reduces to
\begin{equation}
    \diff (\delta \boldsymbol{Q}[\xi] - \xi \cdot \boldsymbol{\theta}(\phi,\delta \phi)) = 0.
\end{equation} 

\begin{figure}[t]
    \centering
    \includegraphics[width=.5\linewidth]{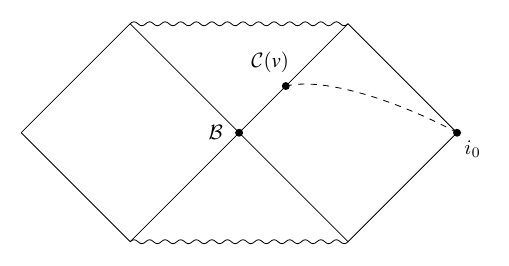}   
    \caption{The Penrose diagram of an asymptotically flat black hole. The non-stationary comparison first law relates the variations of the mass and angular momentum of the spacetime to the variation of the dynamical entropy evaluated on an arbitrary horizon cross-section.}
    \label{fig:comparison_1st_law}
\end{figure}

To derive the comparison first law for an arbitrary horizon cross-section $\mathcal{C}$, we integrate this equation over a codimension$-1$ spatial hypersurface $\S$ bounded by an arbitrary horizon cross-section $\mathcal{C}$ and the cross-section $\mc{S}_\infty$ at spatial infinity $i_0$, as shown in Figure \ref{fig:comparison_1st_law}. According to Stokes' theorem we obtain
\begin{equation}
    \int_{\mc{S}_\infty} (\delta \boldsymbol{Q}[\xi] - \xi \cdot \boldsymbol{\theta}(\phi,\delta \phi)) 
    = \int_{\mc{C}} (\delta \boldsymbol{Q}[\xi] - \xi \cdot \boldsymbol{\theta}(\phi,\delta \phi)).
\end{equation}
For a stationary, axisymmetric black hole, the horizon Killing vector field can be normalized as
\begin{equation}
    \xi^a = (\partial_t)^a + \O_{\mc{H}} (\partial_\vartheta)^a,
\end{equation}
where $(\partial_t)^a$ generates time translations at spatial infinity, $(\partial_\vartheta)^a$ denotes the axial Killing vector, and $\O_\mc{H}$ is the angular velocity of the horizon. So that we identify the integral at spatial infinity with the variation of the mass and angular momentum of the spacetime\cite{Iyer:1994ys}
\begin{equation}
    \int_{\mc{S}_\infty} (\delta \boldsymbol{Q}[\xi] - \xi \cdot \boldsymbol{\theta}(\phi,\delta \phi)) = \delta M - \O_\mc{H} \delta J.
\end{equation}
And we would like to define the dynamical black hole entropy as the “improved” Noether charge\cite{Hollands:2024vbe,Visser:2024pwz}
\begin{equation}\label{dyn identification}
    \frac{\kappa}{2 \pi} \delta S_{\text{dyn}}[\mc{C}] = \int_\mc{C} (\delta \boldsymbol{Q}[\xi] - \xi \cdot \boldsymbol{\theta}(\phi,\delta\phi)),
\end{equation}
where $\k$ denotes the surface gravity of the black hole. If such definition is well posed, the non-stationary comparison first law for an arbitrary horizon cross-section reads
\begin{equation}
    \frac{\kappa}{2 \pi} \delta S_{\text{dyn}}[\mc{C}] = \delta M - \O_\mc{H} \delta J.
\end{equation}
For source-free perturbations the comparison first law holds for any horizon cross-sections, which indicates that the dynamical black hole entropy is a constant at first order in perturbation theory. To study the non-trivial time evolution of $S_{\text{dyn}}$ we may open an external stress-energy $\d T_{ab}$ in the first order perturbation. When an external stress-energy $\d T_{ab}$ is present, the fundamental identity (\ref{stationary background}) becomes\cite{Hollands:2024vbe}
\begin{equation}\label{sourced fi}
    \diff (\delta \boldsymbol{Q}[\xi] - \xi \cdot \boldsymbol{\theta}(\phi,\delta \phi)) = - \xi^a \d \boldsymbol{C}_a
\end{equation}
where $\d \boldsymbol{C}_a$ is given by
\begin{equation}
    \d C_{a a_1 \cdots a_{n-1}}
    = \d T_{ae} {\e^{e}}_{a_1 \cdots a_{n-1}}.
\end{equation}
Integrating (\ref{sourced fi}) over the portion of the horizon between two arbitrary cross-sections $\mc{C}_1$ and $\mc{C}_2$ returns the physical process first law
\begin{equation}
    \frac{\k}{2 \pi} \Delta \d S_{\text{dyn}}
    = \Delta \d M - \O_{\mc{H}} \Delta \d J,
\end{equation}
where the change of the mass and angular momentum of the black hole $\Delta \d M - \O_{\mc{H}} \Delta \d J$ is related to the matter Killing energy flux as follows\cite{Poisson:2009pwt,Gao:2001ut}
\begin{equation}
    \Delta \d M - \O_{\mc{H}} \Delta \d J
    = \int_{v_1}^{v_2} \diff v \int_{\mc{C}(v)} \diff A \ \d T_{ab} \xi^a k^b.
\end{equation}
It was shown that the identification (\ref{dyn identification}) can always be established for first order perturbations of a stationary black hole, since in that case there exists a quantity $\boldsymbol{B}_\mc{H}$ defined on the black hole horizon such that\cite{Hollands:2024vbe,Visser:2024pwz}
\begin{equation}
    \boldsymbol{\theta} \overset{\mc{H}^+}{=} \delta \boldsymbol{B}_\mc{H},\quad \text{and}  \quad
    \boldsymbol{B}_\mc{H}\overset{\mc{H}^+}{=}0.
\end{equation}
And the dynamical black hole entropy $S_{\text{dyn}}$ valid to leading order in perturbation theory is defined by
\begin{equation}
    S_{\text{dyn}}[\mc{C}] = \frac{2 \pi}{\kappa} \int_\mc{C}( \boldsymbol{Q}[\xi] - \xi \cdot \boldsymbol{B}_\mc{H}).
\end{equation}

Finally let's review the dynamical black hole entropy formula in general relativity. The Lagrangian form for general relativity with a cosmological constant $\Lambda$ is
\begin{equation}
    \boldsymbol{L} = \frac{1}{16 \pi G} (R - 2 \L)\boldsymbol{\epsilon},
\end{equation}
where $\boldsymbol{\epsilon}$ is the volume form. And in the following discussion we use the notation
\begin{equation}
    \boldsymbol{\epsilon}_{a_1 \cdots a_p} = \epsilon_{a_1 \cdots a_p a_{p+1} \cdots a_D}.
\end{equation}
For example, $\boldsymbol{\epsilon}_{a}$ denotes the volume form with the first index displayed and the other indices suppressed. The symplectic potential $\boldsymbol{\theta}(\phi, \delta \phi)$ and the Noether charge $\boldsymbol{Q}[\xi]$ of this Lagrangian are\cite{Iyer:1994ys}
\begin{equation}
    \boldsymbol{\theta}(\phi, \delta \phi) = \frac{1}{16 \pi G} g^{ab} g^{cd} (\nabla_c \delta g_{bd} - \nabla_b \delta g_{cd})\boldsymbol{\epsilon}_a,
\end{equation}
\begin{equation}
    \boldsymbol{Q} [\xi] = - \frac{1}{16 \pi G} \boldsymbol{\epsilon}_{ab}\nabla^a \xi^b.
\end{equation}
For a given horizon cross-section $\mc{C}$, we have
\begin{equation}
    \boldsymbol{\e} \overset{\mc{H}^+}{=} k \wedge l \wedge \boldsymbol{\e}_\mc{C},
\end{equation}
where $k^a$ and $l^a$ denote the (future directed) outgoing and ingoing null normal to the cross-section respectively, and they are normalized as $k^a l_a = -1$. $\boldsymbol{\e}_\mc{C}$ represents the codimension$-2$ form of the cross-section. So that
\begin{equation}
    \boldsymbol{Q}  [\xi]  \overset{\mc{C}}{=} - \frac{1}{16 \pi G} \boldsymbol{\epsilon}_{\mc{C}}(k_a l_b - l_a k_b)\nabla^a \xi^b
    \overset{\mc{C}}{=} \frac{\kappa}{8 \pi G} \boldsymbol{\epsilon}_{\mc{C}}.
\end{equation}
And with suitable gauge conditions on perturbations one can prove that\cite{Hollands:2024vbe,Visser:2024pwz}
\begin{equation}
    \xi \cdot \boldsymbol{\theta}(\phi, \delta \phi) \overset{\mc{C}}{=}\frac{1}{8 \pi G}  \boldsymbol{\epsilon}_{\mc{C}} \ \kappa v \ \delta \theta_v
    \overset{\mc{C}}{=} \delta \left(\frac{1}{8 \pi G}\boldsymbol{\epsilon}_{\mc{C}}  \kappa v \theta_v\right),
\end{equation}
where $v$ is the affine null parameter on the event horizon. Therefore the dynamical black hole entropy in general relativity reads
\begin{equation}
    S_{\text{dyn}}[\mc{C}(v) ] = \frac{1}{4G} \int_\mc{C} \left(1 - v  \theta_v\right) \boldsymbol{\epsilon}_{\mc{C}}= \frac{1}{4G} (1 - v \partial_v) A[\mc{C}(v)].
\end{equation}

%%%%%%%%%%%%%%%%%%%%%%%%%%%%%%%%%%%%%%%%%%%%%%%%%%%%%%%%%

\section{Geometric setup} \label{sec3}

In this section we introduce the geometric background of the stationary asymptotically de Sitter black hole spacetime, and impose gauge conditions on non-stationary perturbations.

\begin{figure}[t!]
\centering
\includegraphics[width=.8\linewidth]{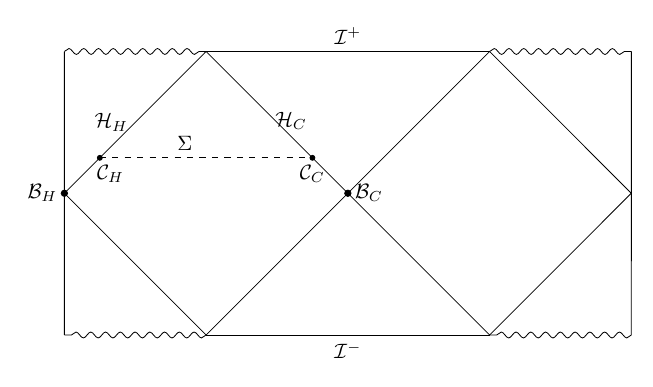}
\caption{\small The Penrose diagram of the asymptotically de Sitter black hole spacetime. The non-stationary comparison first law for an asymptotically de Sitter black hole relates the variation of the dynamical entropy on an arbitrary black hole horizon cross-section $\mc{C}_H$ to that on an arbitrary cosmological horizon cross-section $\mc{C}_C$.
}
\label{fig:dS_BH}  
\end{figure}

Consider an electrically neutral asymptotically de Sitter stationary black hole spacetime as shown in Figure \ref{fig:dS_BH}. We shall adopt the definition of \cite{Gibbons:1977mu} to define the event horizon of the spacetime as the boundary of the past of $\lambda$, i.e., $\dot{I}(\lambda)$, where $\lambda$ is a future inextensible timelike curve representing an observer's world line. We assume that both the black hole event horizon $\mc{H}_H$ and the cosmological event horizon $\mc{H}_C$ possess the structure of bifurcate Killing horizons. And we are mainly interested in the parts of horizons that lie to the future of the bifurcation surfaces $\mc{B}_H$ and $\mc{B}_C$.  We denote the affinely parameterized null generators of the black hole event horizon and the cosmological event horizon by $k_H^a$ and $k_C^a$ respectively, and set the affine parameters to be $0$ at the bifurcation surfaces. The Killing field that is normal to the cosmological event horizon is denoted by $\xi^a$. Then the Killing vector $\hat{\xi}^a$ which coincides with the generators of the black hole event horizon can be expressed in the form\cite{Gibbons:1977mu}
\begin{equation}\label{BH Killing}
    \hat{\xi}^a = \xi^a + \O_H \varphi^a,
\end{equation}
where $\O_H$ is the angular velocity of the black hole horizon relative to the cosmological horizon, and $\varphi^a$ is the axial Killing vector whose orbits are closed curves with parameter length $2\pi$. We denote the surface gravities of the black hole horizon and the cosmological horizon by $\k_H$ and $\k_C$ respectively, then
\begin{equation}
    \hat{\xi}^a \overset{\mc{H}_H}{=} \kappa_H v_H k_H^a, \quad  
    \xi^a \overset{\mc{H}_C}{=} \kappa_C v_C k_C^a,
\end{equation}
where $v_H$ and $v_C$ are affine parameters along the black hole horizon and the cosmological horizon respectively.

We would like to perturb the stationary asymptotically de Sitter background $g \to g + \delta g$ and the matter fields $\phi \to \phi + \d \phi$ within it to study the first law, where $\d g, \ \d \phi \sim \mc{O}(\e)$ are of first order in perturbation theory. We are mainly interested in non-stationary perturbations, which are defined as
\begin{equation}
    \delta (\mc{L}_\xi g) \ne 0, \ \delta (\mc{L}_\xi \phi) \ne 0.
\end{equation}
When comparing two slightly different spacetimes there exists certain freedom in which spacetime points are chosen to correspond. In order to simplify the derivation of the first law, we impose the following gauge conditions on perturbations:
\begin{itemize}
    \item The black hole event horizon and the cosmological event horizon of the perturbed spacetime are identified with those in the stationary background.

    \item We take the Killing vector $\xi^a$ that generates the cosmological event horizon and the axial Killing vector $\varphi^a$ to be fixed under the variation, i.e., 
    \begin{equation}
        \delta \xi^a = 0, \quad \delta \varphi^a = 0.
    \end{equation}

    \item We impose the following two sets of conditions on $\d g_{ab}$ at the black hole event horizon and the cosmological event horizon:
    \begin{equation}\label{gauge condition C}
        \xi^a \d g_{ab} \overset{\mc{H}_C}{=}0, \quad 
        \nabla_a (\xi^b \xi^c \d g_{bc}) \overset{\mc{H}_C}{=}0.
    \end{equation}
    \begin{equation}\label{gauge condition H}
        \hat\xi^a \d g_{ab} \overset{\mc{H}_H}{=}0, \quad 
        \nabla_a (\hat\xi^b \hat\xi^c \d g_{bc}) \overset{\mc{H}_H}{=}0.
    \end{equation}  
    Condition (\ref{gauge condition C}) requires that $\xi^a$ remains the null normal to the cosmological event horizon, and the surface gravity $\k_C$ is fixed under the variation\cite{Hollands:2024vbe}. The same connotations hold for the condition (\ref{gauge condition H}).
\end{itemize}

We emphasize these conditions do not mean that $\hat\xi^a$ should be fixed under the perturbations. For perturbations that change the horizon angular velocity $\d \O_H \ne 0$, The Killing field $\hat\xi^a$ varies, i.e.,
\begin{equation}
    \d \hat\xi^a = \d \O_H \varphi^a.
\end{equation}

%%%%%%%%%%%%%%%%%%%%%%%%%%%%%%%%%%%%%%%%%%%%%%%%%%%%%%%%%

\section{The non-stationary first law for de Sitter horizons} \label{sec4}

In this section we utilize the covariant phase space formalism to derive both the non-stationary comparison first law for de Sitter black holes, and the non-stationary physical process first law for cosmological event horizons.

\subsection{The comparison first law}
In the presence of minimally coupled external matter fields, we divide the Lagrangian $\boldsymbol{L}$ of the system into the pure gravitational part $\boldsymbol{L}_g(g) = \frac{1}{16 \pi G} (R - 2 \L) \boldsymbol{\e}$ depending only on the metric, and the matter part $\boldsymbol{L}_m(g,\psi)$ depending on both the metric and the external matter fields
\begin{equation}
    \boldsymbol{L}(g,\psi) = \boldsymbol{L}_g(g) + \boldsymbol{L}_m(g,\psi).
\end{equation}
Under a first-order variation of the dynamical fields we obtain
\begin{equation}
    \begin{split}
        \delta \boldsymbol{L}_g(g) &= \frac{1}{2} \boldsymbol{E}_{ab} \delta g^{ab} + \diff \boldsymbol{\theta}_g(g ,\delta g), \\
    \delta \boldsymbol{L}_m(g,\psi) &= -\frac{1}{2} \boldsymbol{T}_{ab} \delta g^{ab} + \boldsymbol{E}_m \delta \psi + \diff \boldsymbol{\theta}_m(g,\psi ,\delta \psi),
    \end{split}
\end{equation}
where $E_{ab} =\frac{1}{8 \pi G}(R_{ab} - \frac{1}{2} R g_{ab} + \L g_{ab})$ is the gravitational field equation, $T_{ab}$ is the stress-energy tensor of the external matter fields, and $E_m$ is the equation of motion for $\psi$. Next we decompose the relevant forms $\boldsymbol{\theta}$, $\boldsymbol{\omega}$ and $\boldsymbol{J}$ of the Lagrangian $\boldsymbol{L}(g,\psi)$ into gravitational parts and matter parts. For perturbations that left the Killing vector $\xi^a$ invariant, the symplectic $(n-1)$-form $\boldsymbol{\o}(\phi;\d \phi, \mc{L}_{\xi}\phi)$ derived from the Lagrangian of the total system $\boldsymbol{L}(g,\psi)$ satisfies\cite{Iyer:1996ky}
\begin{equation}
    \boldsymbol{\o}(\phi;\d \phi, \mc{L}_{\xi}\phi)
    = \diff (\delta \boldsymbol{Q}_m[\xi] - \xi \cdot \boldsymbol{\theta}_m(\phi,\delta \phi))
    - \delta (T^{ab} \xi_b \boldsymbol{\epsilon_a}) - \frac{1}{2}\xi \cdot \boldsymbol{T}_{ab} \delta g^{ab}.
\end{equation}
On the other hand, since the background field equations and constraint equations of the system are satisfied, the fundamental identity (\ref{fundamental identity}) indicates that
\begin{equation}
    \begin{split}
        \boldsymbol{\o}(\phi;\d \phi, \mc{L}_{\xi}\phi)
    &= \diff (\delta \boldsymbol{Q}[\xi] - \xi \cdot \boldsymbol{\theta}(\phi,\delta \phi)) \\
    &= \diff (\delta \boldsymbol{Q}_g[\xi] - \xi \cdot \boldsymbol{\theta}_g(g,\delta g))
    + \diff (\delta \boldsymbol{Q}_m[\xi] - \xi \cdot \boldsymbol{\theta}_m(\phi,\delta \phi)).
    \end{split}
\end{equation}
Thus the Noether charge of the gravitational part $\boldsymbol{Q}_g$ satisfies
\begin{equation}\label{fundamental identity for Qg}
    \diff (\delta \boldsymbol{Q}_g[\xi] - \xi \cdot \boldsymbol{\theta}_g(g,\delta g)) =  - \delta (T^{ab} \xi_b \boldsymbol{\epsilon_a}) - \frac{1}{2}\xi \cdot \boldsymbol{T}_{ab} \delta g^{ab}.
\end{equation}
Next we integrate this formula on a spacelike hypersurface $\Sigma$ bounded by an arbitrary black hole horizon cross-section $\mc{C}_H$ and an arbitrary cosmological horizon cross-section $\mc{C}_C$, as shown in Figure \ref{fig:dS_BH}. According to the Stokes' theorem, the result is
\begin{equation}
    \int_\Sigma \left[ \delta (T^{ab} \xi_b \boldsymbol{\epsilon_a}) + \frac{1}{2}\xi \cdot \boldsymbol{T}_{ab} \delta g^{ab} \right]  = 
    - \int_{\mc{C}_C}(\delta \boldsymbol{Q}_g[\xi] - \xi \cdot \boldsymbol{\theta}_g(g,\delta g))
    - \int_{\mc{C}_H} (\delta \boldsymbol{Q}_g[\xi] - \xi \cdot \boldsymbol{\theta}_g(g,\delta g)).
\end{equation}
On the right hand side, the integral over the cosmological event horizon is identified as the variation of the dynamical entropy evaluated on $\mc{C}_C$
\begin{equation}\label{dynamical entropy}
    \int_{\mc{C}_C}(\delta \boldsymbol{Q}_g[\xi] - \xi \cdot \boldsymbol{\theta}_g(g,\delta g)) = \frac{\k_C}{2 \pi} \delta S_\text{dyn}[\mc{C}_C],
\end{equation}
where
\begin{equation}
    S_\text{dyn}[\mc{C}_C] = \frac{2 \pi}{\k_C} \int_{\mc{C}_C} (\boldsymbol{Q}_g[\xi] - \xi \cdot \boldsymbol{B}_\mc{H}(g,\delta g))= \frac{1}{4G} \left( 1 - v_C \frac{\diff}{\diff v_C} \right) A[\mc{C}_C].
\end{equation}
According to (\ref{BH Killing}), the integral over the black hole event horizon can be manipulated as follows
\begin{equation}
\begin{split}
    &\int_{\mc{C}_H}(\delta \boldsymbol{Q}_g[\xi] - \xi \cdot \boldsymbol{\theta}_g(g,\delta g)) \\
    = &\int_{\mc{C}_H}[\delta (\boldsymbol{Q}_g[\hat\xi] - \O_H \boldsymbol{Q}_g[\varphi]) - \hat\xi \cdot \boldsymbol{\theta}_g(g,\delta g)] \\
    = &\int_{\mc{C}_H}[\delta_\phi \boldsymbol{Q}_g[\hat\xi] + \boldsymbol{Q}_g[\d\hat\xi] - \d\O_H \boldsymbol{Q}_g[\varphi] - \O_H \d \boldsymbol{Q}_g[\varphi] - \hat\xi \cdot \boldsymbol{\theta}_g(g,\delta g)] \\
    =&\int_{\mc{C}_H}(\delta_\phi \boldsymbol{Q}_g[\hat\xi] - \hat\xi \cdot \boldsymbol{\theta}_g(g,\delta g)) 
    - \O_H \int_{\mc{C}_H} \delta \boldsymbol{Q}_g[\varphi],
\end{split}
\end{equation}
where $\d_\phi$ denotes the variation that only acts on the dynamical fields. And in the second equality we have used $\varphi \cdot \boldsymbol{\theta}_g$ vanishes when pull it back on $\mc{C}_H$ since $\varphi$ is parallel to $\mc{C}_H$. In the third equality we have used $\d \hat\xi^a = \d \O_H \varphi^a$. The first integral in the last line is identified as the variation of the dynamical entropy of the black hole horizon
\begin{equation}
    \int_{\mc{C}_H}(\delta_\phi \boldsymbol{Q}_g[\hat\xi] - \hat\xi \cdot \boldsymbol{\theta}_g(g,\delta g)) 
    = \frac{\k_H}{2\pi} \d S_\text{dyn}[\mc{C}_H],
\end{equation}
where 
\begin{equation}
    S_\text{dyn}[\mc{C}_H] = \frac{1}{4G} \left( 1 - v_H \frac{\diff}{\diff v_H} \right) A[\mc{C}_H].
\end{equation}
And we make the identification such that
\begin{equation}
   \d J_H = \int_{\mc{C}_H} \d \boldsymbol{Q}_g[\varphi]. 
\end{equation}
Since for a stationary de Sitter black hole, $J_H = \int_{\mc{C}_H} \boldsymbol{Q}_g[\varphi] = \frac{1}{16 \pi} \int_{\mc{C}_H} \boldsymbol{\epsilon}_{ab} \nabla^a \varphi^b$ recovers the definition of the angular momentum of the black hole\cite{Gibbons:1977mu}.
Therefore the non-stationary comparison first law between an arbitrary black hole horizon cross-section $\mc{C}_H$ and an arbitrary cosmological horizon cross-section $\mc{C}_C$ reads
\begin{equation}
    \int_\Sigma \left[ \delta (T^{ab} \xi_b \boldsymbol{\epsilon_a}) + \frac{1}{2}\xi \cdot \boldsymbol{T}_{ab} \delta g^{ab} \right]
    = - \frac{\k_C}{2\pi} \delta S_\text{dyn}[\mc{C}_C] - \frac{\k_H}{2\pi} \delta S_\text{dyn}[\mc{C}_H] - \O_H \delta J_H.
\end{equation}
If external matter fields are absent in the stationary background $T_{ab} = 0$, we obtain
\begin{equation}
    \int_\Sigma  \d T_{ab} \xi^a \diff \S^b
    = - \frac{\k_C}{2\pi} \delta S_\text{dyn}[\mc{C}_C] - \frac{\k_H}{2\pi} \delta S_\text{dyn}[\mc{C}_H] - \O_H \delta J_H.
\end{equation}
The non-stationary first law of event horizons for arbitrary horizon cross-sections still takes the form of (\ref{GHfirstlaw}), while the Bekenstein-Hawking entropy should be replaced by the dynamical entropy in general relativity.

\subsection{The physical process first law}

\begin{figure}[t!]
\centering
\includegraphics[width=.4\linewidth]{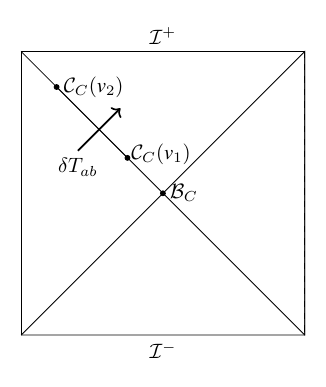}
\caption{\small The physical process first law for the de Sitter horizon. An external matter source denoted by $\d T_{ab}$ passes through the cosmological event horizon between two arbitrary cross-sections $\mc{C}_C(v_1)$ and $\mc{C}_C(v_2)$. The physical process first law computes the entropy change between $\mc{C}_C(v_1)$ and $\mc{C}_C(v_2)$ in this process.}
\label{fig:dS_PPFL}  
\end{figure}

The derivation of the physical process first law for de Sitter horizons is very similar to the case of black hole physics\cite{Visser:2024pwz,Hollands:2024vbe}. We integrate (\ref{fundamental identity for Qg}) on the cosmological event horizon between two arbitrary cross-sections $\mc{C}_C(v_1)$ and $\mc{C}_C(v_2)$, as shown in Figure \ref{fig:dS_PPFL}. Since $\xi^a$ is tangent to the cosmological event horizon $\mc{H}_C$, the second term on the right hand side of (\ref{fundamental identity for Qg}) vanishes when pull it back on the horizon. On the other hand, the null-null component of the stress-energy tensor $T_{vv}$ vanishes in the stationary background\cite{Poisson:2009pwt}. So that the first term on the right hand side $\d (T^{ab}\xi_b \boldsymbol{\e}_a)$ reduces to $\d T_{ab} \xi^a k^b \boldsymbol{\e}_C$. And we obtain
\begin{equation}
        \left(\int_{\mc{C}_C(v_2)} - \int_{\mc{C}_C(v_1)} \right)(\delta \boldsymbol{Q}_g[\xi] - \xi \cdot \boldsymbol{\theta}_g(g,\delta g))
        = \int_{v_1}^{v_2} \diff v \int_{\mc{C}_C(v)} \diff A \ \d T_{ab} \xi^a k^b.
\end{equation}
Recalling the definition of the dynamical entropy (\ref{dynamical entropy}) we get\footnote{It's also worth pointing out that the non-stationary physical process first law in general relativity can be obtained by integrating the linearized Raychaudhuri equation on the horizon\cite{Rignon-Bret:2023fjq,Visser:2024pwz,Kong:2024sqc}.}
\begin{equation}
    \frac{\k_C}{2 \pi}\Delta \d S_{\text{dyn}} = \int_{v_1}^{v_2} \diff v \int_{\mc{C}_C(v)} \diff A \ \d T_{ab} \xi^a k^b.
\end{equation}
If the matter fields falling through the cosmological event horizon satisfies null energy condition $\d T_{ab} k^a k^b \ge 0$, then the physical process first law suggests that the linearized second law is obeyed for perturbations sourced by external matter fields. Notice that the change of the mass of the matter fields inside the cosmological event horizon is related to the matter Killing energy flux by\cite{Bousso:2002fq,Galante:2023uyf} 
\begin{equation}
    \Delta \d M = - \int_{v_1}^{v_2} \diff v \int_{\mc{C}_C(v)} \diff A \ \d T_{ab} \xi^a k^b,
\end{equation}
As a result, the non-stationary physical process first law for de Sitter horizons is described by
\begin{equation}
    \Delta \d M = - \frac{\k_C}{2 \pi}\Delta \d S_{\text{dyn}}.
\end{equation}

 %%%%%%%%%%%%%%%%%%%%%%%%%%%%%%%%%%%%%%%%%%%%%%%%%%%%%%%%%

\section{Conclusion and discussion} \label{sec5}

 We have lifted the construction of the dynamical entropy to cosmological event horizons in asymptotically de Sitter spacetimes. By applying the Noether charge method to non-stationary perturbations of a stationary de Sitter black hole spacetime, we have demonstrated that dynamical entropy satisfies both
 the non-stationary comparison first law and the non-stationary physical process first law.
 All of these further encourage us to extend the dynamical entropy to more general horizons\cite{Jacobson:2003wv,Yan:2024gbz}, such as expansion and shear free horizons. 

 %%%%%%%%%%%%%%%%%%%%%%%%%%%%%%%%%%%%%%%%%%%%%%%%%%%%%%%%%

\noindent \noindent
\section*{Acknowledgments}

I am grateful to Delong Kong for helpful discussions. I also thank Zihan Yan for useful correspondence. This work is partly supported by the National Key Research and Development Program of China with Grant No. 2021YFC2203001 as well as the National Natural Science Foundation of China with Grants No. 12075026 and No. 12361141825.

%%%%%%%%%%%%%%%%%%%%%%%%%%%%%%%%%%%%%%%%%%%%%%%%%%%%%%%%%

 \bibliographystyle{JHEP}
 \bibliography{dSrefs}

\end{document}